\documentstyle[12pt,epsfig]{article}                                             
                                             
\newlength{\dinwidth}                                             
\newlength{\dinmargin}                                             
\def\lapproxeq{\lower .7ex\hbox{$\;\stackrel{\textstyle                                             
<}{\sim}\;$}}                                             
\def\gapproxeq{\lower .7ex\hbox{$\;\stackrel{\textstyle                                             
>}{\sim}\;$}}                                             
\def\be{\begin{equation}}                                             
\def\ee{\end{equation}}                                             
\def\bea{\begin{eqnarray}}                                             
\def\eea{\end{eqnarray}}                                             
                                             
\def\funp{{I\!\!P}}             
                                             
\begin{document}                                             
\begin{flushright}                                             
DTP/00/08 \\                                             
February 2000 \\                                             
\end{flushright}                                             
                                             
\vspace*{2cm}                                             
                                             
\begin{center}                                             
{\Large \bf Can the Higgs be seen in rapidity gap events} \\    
    
\vspace*{0.5cm}    
{\Large \bf at the Tevatron or the LHC?}                                             
                                             
\vspace*{1cm}                                             
V.A. Khoze$^a$, A.D. Martin$^a$ and M.G. Ryskin$^{a,b}$ \\                                             
                                            
\vspace*{0.5cm}                                             
$^a$ Department of Physics, University of Durham, Durham, DH1 3LE \\                                            
$^b$ Petersburg Nuclear Physics Institute, Gatchina, St.~Petersburg, 188350, Russia     
\end{center}                                             
                                             
                                             
\begin{abstract}                                             
Double diffractive Higgs production at $pp$ (or $\bar{p}p$) colliders continues to attract   
attention as a potential signal in the search for the boson.  We present improved perturbative  
QCD estimates of the event rates for both the exclusive and inclusive double diffractive  
Higgs processes, paying particular attention to the survival probability of the rapidity gaps.   
We find that the major uncertainty is in the prediction for the survival probability associated  
with soft rescattering.  We show that an analogous process, the double diffractive production  
of a pair of jets with large values of $E_T$, has an event rate which makes it accessible at the  
Tevatron.  Observation of this process can therefore be used as a luminosity monitor for two- 
gluon exchange processes, such as the production of a Higgs boson with rapidity gaps on  
either side.  
\end{abstract}                                   
    
\section{Introduction}  
  
A central problem in particle physics is to find a good signal with which to identify the Higgs     
boson at the present and forthcoming hadron colliders, the Tevatron and the LHC.  This has   
become more important since it appears likely that the Higgs boson will be beyond the reach  
of LEP2.  One possibility which, at first sight, looks attractive is to select events with a large  
rapidity gap on either side, where the conventional background is relatively low \cite{DKT}-- 
\cite{BL}.  From an experimental point of   
view the exclusive signal looks particularly promising   
\be    
\label{eq:a1}    
p + p \; \rightarrow \; p + {\rm gap} + H + {\rm gap} + p,    
\ee    
and similarly for $p\bar{p}$ events.  For an exclusive process we have the possibility of good   
experimental resolution on the Higgs boson mass, $M_H$, whereas in an inelastic collision   
the event rate is higher but the large multiplicity of secondary particles poses an additional  
problem in identifying the Higgs boson. The main question is whether the production rate of  
Higgs events with rapidity gaps is large enough.   
  
The cross section for double diffractive rapidity gap Higgs production can be   
estimated using perturbative QCD but unfortunately it is found \cite{KMR} that it is strongly   
suppressed by rescattering and QCD radiative effects.  Despite this, very optimistic  
predictions of the exclusive event rate persist, and are frequently cited in experimental  
proposals.  The purpose of this paper is to improve the reliability of the perturbative QCD  
predictions by going beyond double $\log$ accuracy so as to give believable   
estimates of the event rate and to settle the present ambiguities.  We will see that this  
improvement will lead to some enhancement of the event rate as compared to our previous  
estimates \cite{KMR,MRK}.  Moreover, since the basic QCD mechanism for Higgs  
production is the same as that for the double diffractive central production of a pair of large  
$E_T$ jets, we also estimate the event rate for the latter process (which has a larger cross   
section) so that it may be used as a pomeron-pomeron luminosity monitor for rapidity gap   
Higgs production.  Indeed such dijet data are already accessible at the Tevatron, see, for  
example, \cite{TER} and references therein.  Thus we have a valuable check on the QCD  
predictions for process (\ref{eq:a1}).  
  
As mentioned above, the cross section for Higgs production via the exclusive process   
(\ref{eq:a1}) is suppressed by the small survival probability of the rapidity gaps.  The survival   
probability $w$ is given by the product of two factors\footnote{We do not discuss the  
multiple (or \lq\lq pile-up\rq\rq) interactions of high luminosities.}  
\be  
\label{eq:b1}  
w \; = \; S_{\rm spec}^2 \: T^2.  
\ee  
First, the gaps may be filled by soft parton rescattering and, second, by QCD bremsstrahlung  
from the two fast coloured gluons which annihilate into the Higgs boson, see Fig.~1.  The   
probability $S_{\rm spec}^2$ not to have any extra soft rescattering was estimated in   
\cite{DKS,BJ,LE,GLM} to be about $S_{\rm spec}^2 \simeq 0.1$, up to a factor of 2.  This   
suppression agrees with the simple phenomenological estimate \cite{RR}  
\be  
\label{eq:c2}  
S_{\rm spec}^2 \; = \; \langle e^{- \Omega (\rho_T)} \rangle \; \simeq \; \left (1 \: - \: \frac{2   
\sigma^D}{\sigma_{\rm tot}} \right )^2,  
\ee  
where $\Omega$ is the opacity (or optical density) of the proton and $\langle \ldots \rangle$  
indicates the appropriate average   
over the impact parameter $\rho_T$; $\sigma^D$ is the sum of the elastic and diffractive   
dissociation cross sections and $\sigma_{\rm tot}$ is the $pp$ (or $p\bar{p}$) total cross   
section.  We will return to discuss the determination of $S_{\rm spec}^2$ in detail in Section  
5.  
  
The factor $T^2$ in (\ref{eq:b1}) is the probability not to radiate gluons in the hard   
subprocess $gg \rightarrow H$, and is incorporated in the perturbative QCD calculation of the   
exclusive amplitude.  To suppress the QCD radiation we have to screen the colour of the   
annihilating gluons by an additional $t$-channel gluon, as in Fig.~1.  The most optimistic   
scenario is to assume that this gluon, which screens the colour,     
does not couple to the Higgs and that it has small virtuality $Q_T^2$ to enhance the     
probability of screening via a large value of $\alpha_S$.  However there is a probability of     
relatively hard gluon emission coming from distance scales $\lambda \gapproxeq 1/M_H$,  
but shorter than the characteristic transverse size ($\sim 1/Q_T$) at which the colour flow is     
screened.  Such gluons fill up the rapidity gaps.  It is found \cite{KMR} that the typical     
values of $Q_T$ of the screening \lq soft\rq~gluon are indeed much smaller than $M_H$, but   
yet are sufficiently large for perturbative QCD to be applicable.    
  
Since there has been much recent interest in the double diffractive Higgs signal, both for  
experiments at the Tevatron and the LHC (see, for example, \cite{NEW}), it is timely to  
reassess the estimate of the event rate.  
    
\section{Double diffractive exclusive Higgs production}  
  
In Ref.~\cite{KMR} the production amplitude for the exclusive process (\ref{eq:a1}), derived   
from perturbative QCD, was given by  
\be    
\label{eq:a2}    
{\cal M} (pp \rightarrow p + H + p) \; = \; A \pi^3 \: \int \: \frac{dQ_T^2}{Q_T^4} \: e^{- S     
(Q_T^2, M_H^2)} \: f (x_1, Q_T^2) \: f (x_2, Q_T^2),    
\ee    
where $f (x, Q_T^2)$ is the unintegrated gluon density of the proton and $A$ is a factor     
associated with the $gg \rightarrow H$ vertex\footnote{We assume that $M_H$ is below  
$m_t$ and hence is far from the $t\bar{t}$ threshold.}    
\be    
\label{eq:b2}    
A \; = \; (\sqrt{2} G_F)^{\frac{1}{2}} \: \alpha_S (M_H^2)/3\pi.    
\ee    
The exponential is the conventional double $\log$ Sudakov form factor which is the  
probability {\it not} to emit bremsstrahlung gluons (one of which is shown in Fig.~1a by  
$p_T$) in the interval $Q_T \lapproxeq p_T \lapproxeq M_H/2$.  The upper bound of $p_T$  
is clear, and the lower bound occurs because there is destructive interference of the amplitude  
in which the bremsstrahlung gluon is emitted from a \lq\lq hard\rq\rq~gluon $k_i$ with that in  
which it is emitted from the screening gluon.  That is there is no emission when $\lambda  
\simeq 1/p_T$ is larger than the separation $d \sim 1/Q_T$ of the two $t$-channel gluons in  
the transverse plane, since then they act effectively as a colour-singlet system.  So the  
Sudakov form factor (that is the probability not to have bremsstrahlung) is given by the  
Poisson distribution $\exp (-S)$, where the mean multiplicity of bremsstrahlung is  
\be    
\label{eq:a3}    
S (Q_T^2, M_H^2) \; = \; \int_{Q_T^2}^{M_H^2/4} \: \frac{C_A \alpha_S (p_T^2)}{\pi} \:     
\frac{dE}{E} \: \frac{dp_T^2}{p_T^2}.  
\ee    
Here $E$ and $p_T$ are the energy and transverse momentum of an emitted gluon in the     
Higgs rest frame.    
    
Note that the amplitude (\ref{eq:a2}) for the exclusive process is written for forward outgoing   
protons, that is for a Higgs boson produced with small transverse momentum $q_T$.  Indeed,   
the presence of the proton form factors suppresses large $q_T$ production, and the exclusive   
cross section $\left . d \sigma/dy_{H} \right |_0$ is calculated assuming form factors $\exp  
(bt_i/2)$ at the proton vertices, with $b = 5.5$~GeV$^{-2}$.  Here the notation means that  
the cross section is to be evaluated for Higgs rapidity $y_H = 0$.  
  
Eq.~(\ref{eq:a2}) is the perturbative QCD estimate of the exclusive amplitude to double   
$\log$ accuracy.  We are now able to improve the prediction by (i) including the effect of   
using skewed or off-diagonal gluon distributions and (ii) using a more precise definition of  
the unintegrated gluon distribution.  With these improvements the amplitude (\ref{eq:a2})  
may be rewritten, to single $\log$ accuracy\footnote{The single $\log$ accuracy of  
(\ref{eq:b3}) may be established using $Q_T$-factorization \cite{ITAL}.  The crucial point is  
that in a physical gauge (for example, the planar gauge $A_\mu^a n_\mu = 0$ with the gauge  
vector $n_\mu$ chosen as the longitudinal component of the Higgs 4-momentum) any  
additional gluon which embraces the Higgs boson (that is which connects the upper and lower  
parts of the diagram in Fig.~1a) gives neither a DGLAP or BFKL collinear logarithm.}, in the  
form  
\be  
\label{eq:b3}  
{\cal M} (pp \rightarrow p + H + p) \; = \; A \pi^3 \: \int \: \frac{dQ_T^2}{Q_T^4} \: f_g   
(x_1, x_1^\prime, Q_T^2, M_H^2/4) \: f_g (x_2, x_2^\prime, Q_T^2, M_H^2/4)  
\ee  
where $f_g (x, x^\prime, Q_T^2, M_H^2/4)$ denotes the skewed or off-diagonal  
unintegrated gluon density in the initial proton.  The diagonal density is defined such that the  
probability to find a gluon (with transverse momentum $Q_T$ and momentum fraction $x$  
in the interval $d Q_T^2 dx$) is \linebreak $f_g (dQ_T^2/Q_T^2)(dx/x)$.  These  
unintegrated distributions are the quantities which enter when we apply the   
$Q_T$-factorization theorem \cite{ITAL} to the evaluation of the Feynman diagram of  
Fig.~1a.  The procedure of how to calculate $f_g (x, x, Q_T^2, \mu^2)$ from the  
conventional integrated gluon $g (x, Q_T^2)$ is described in Ref.~\cite{KIMBER}.  Here we  
will use the form proposed by DDT \cite{DDT}   
\be  
\label{eq:c3}  
f_g (x, x, Q_T^2, \mu^2) \; = \; \frac{\partial}{\partial \ln Q_T^2} \: \left [ T (Q_T, \mu) \: xg   
(x, Q_T^2) \right ],  
\ee  
where $T (Q_T, \mu)$ is the survival probability that the gluon with $x, x^\prime = x$ and   
transverse momentum $Q_T$ remains untouched in the evolution up to the hard scale $\mu   
(= M_H/2)$.  $T$ is the result of resumming the virtual $(\propto \delta (1 - z))$   
contributions in the DGLAP evolution equation and is given by \cite{KIMBER}  
\be  
\label{eq:d3}  
T (Q_T, \mu) \; = \; \exp \: \left ( - \int_{Q_T^2}^{\mu^2} \: \frac{\alpha_S (k_t^2)}{2 \pi} \:   
\frac{dk_t^2}{k_t^2} \: \int \: \left [z P_{gg} (z) \: + \: \sum_q \: P_{qg} \right ] \: dz \right ).  
\ee  
The derivative $\partial T/\partial \ln Q_T^2$ in (\ref{eq:c3}) cancels the virtual DGLAP   
term in $\partial (xg)/\partial \ln Q_T^2$.  To be precise the equation for $f_g$ is a little more   
complicated than (\ref{eq:c3}) (see eq.~(3) of \cite{KIMBER}).  However in the relevant  
small   
$x$ and $Q_T \ll M_H$ region, (\ref{eq:c3}) is sufficiently accurate for our purposes.  Note   
that after integrating (\ref{eq:c3}) up to scale $\mu$ we do indeed get back the integrated   
gluon distribution  
\be  
\label{eq:e3}  
\int^{\mu^2} \: f_g (x, x, Q_T^2, \mu^2) \: \frac{dQ_T^2}{Q_T^2} \; = \; T (\mu, \mu) \: xg   
(x, \mu^2) \; = \; xg (x, \mu^2).  
\ee  
To double $\log$ accuracy we see from (\ref{eq:d3}) that $T = \exp (- S)$ with $S$ given by   
(\ref{eq:a3}).  In fact we will work to single $\log$ accuracy and hence it follows from  
(\ref{eq:d3}) that  
\be  
\label{eq:f3}  
T (Q_T, \mu) \; = \; \frac{\alpha_S (Q_T^2)}{\alpha_S (\mu^2)} \: e^{-S}.  
\ee  
We comment on this result in Section 7.  
  
Now we must consider the skewed effect coming from the fact that the screening gluon   
carries a much smaller momentum fraction, that is $x_i^\prime \ll x_i$.  As a consequence we   
have  
\be  
\label{eq:g3}  
f_g (x, x^\prime, Q_T^2, M_H^2/4) \; = \; R_g \: \frac{\partial}{\partial \ln Q_T^2} \: \left   
[\sqrt{T  (Q_T, M_H/2)} \: xg (x, Q_T^2) \right ]  
\ee  
where the $\sqrt{T}$ arises because the survival probability is only relevant to the hard   
gluon\footnote{It was shown explicitly in \cite{MR} that when $x^\prime \ll x$ only the   
self-energy of the hard $x$ gluon contributes to the survival probability to leading $\log$   
accuracy.  Note that the gluon with $x^\prime \simeq 0$ is almost \lq\lq at rest\rq\rq~(that is  
$Q_L \ll Q_T$) and there is no possibility of QCD radiation.}.  The multiplicative factor  
$R_g$ is the ratio of the off-diagonal $x^\prime \ll x$   
integrated distribution to the conventional diagonal one $xg (x, Q_T^2)$.  In terms of the   
Operator Product Expansion both the diagonal and off-diagonal (or skewed) distributions are   
given by the expectation values of the same conformal operators \cite{ER,CHASE,O}.  It was   
shown \cite{SGMR} that for $x \ll 1$ the expectation values for the diagonal and skewed   
distributions are the same.  Hence the skewed distribution is completely determined in terms   
of the conventional diagonal gluon.  Indeed for $x^\prime \ll x \ll 1$ we have   
\cite{SGMR}\footnote{Strictly speaking (\ref{eq:h3}) was only proved for integrated gluons   
\cite{SGMR}.  However it is expected to hold equally well for the unintegrated distribution.}  
\be  
\label{eq:h3}  
R_g \; = \; \frac{2^{2 \lambda + 3}}{\sqrt{\pi}} \: \frac{\Gamma \left ( \lambda + \frac{5}{2}   
\right )}{\Gamma (\lambda + 4)}  
\ee  
where $\lambda$ governs the small $x$ behaviour of the diagonal gluon $xg (x, Q_T^2)   
\propto x^{- \lambda}$.  
  
Note that with the $\sqrt{T}$ in (\ref{eq:g3}) the amplitude (\ref{eq:b3}) contains the same   
Sudakov suppression factor, $\exp (-S)$, as in (\ref{eq:a2}).  However amplitude   
(\ref{eq:b3}) includes two improvements in comparison to the previous perturbative QCD   
form (\ref{eq:a2}).  We now include the $\partial T/\partial \ln Q_T^2$ contribution, see   
(\ref{eq:c3}), and also the single $\log$ part of the skewed effect, that is the factor $R_g$ in   
(\ref{eq:g3}).  Both improvements enhance the exclusive $pp \rightarrow p + H + p$ event   
rate.  This is particularly true at FNAL energies where the $\partial T/\partial \ln Q_T^2$   
dominates the $\partial g/\partial \ln Q_T^2$ contribution in (\ref{eq:c3}).  Moreover the   
enhancement due to $R_g$ is non-negligible.  In the relevant kinematic domain we have   
$R_g \simeq 1.2 (1.4)$ leading to an enhancement factor $R_g^4 \simeq 2 (4)$ at LHC   
(Tevatron) energies.  
  
The values of the double diffractive exclusive Higgs cross section at Tevatron and LHC  
energies that are obtained from (\ref{eq:b3}) are presented in Section 4.  We emphasize that  
this perturbative QCD calculation is based on the unintegrated gluon distribution $f_g$  
obtained from (\ref{eq:g3}).  It has been checked to be realistic in the sense that it gives  
reasonable cross sections for diffractive vector meson $(\rho, J/\psi, \Upsilon)$ production  
\cite{MRT4} and for large $q_t$ prompt photon hadroproduction at the Tevatron energy  
\cite{KIMBER}.  We will see the importance of this comment in Section 7.  
  
\section{Double diffractive inclusive Higgs production}  
  
The cross section for the inclusive process     
\be    
\label{eq:a4}    
p + p \; \rightarrow \; X + {\rm gap} + H + {\rm gap} + Y    
\ee    
is much larger than for the exclusive process (\ref{eq:a1}), see Ref.~\cite{KMR}.  Here the  
initial protons may be broken up and so the transverse momentum of the Higgs is no longer  
limited by the proton form factor, and hence the Sudakov suppression is weaker.   
Unfortunately we can no longer achieve single $\log$ accuracy.  The momenta transferred,  
$t_i = (Q - k_i)^2$, are large and hence we cannot express the \lq\lq blobs\rq\rq~in Fig.~1 in  
terms of the gluon density of the proton.  At present, the corresponding skewed (large $t_i$)  
unintegrated gluon distribution is not known.  So we must use the (BFKL-type) non-forward  
amplitude.  On the other hand for inclusive production we allow emission in a larger part of  
the phase space and so the QCD suppression is weaker, and the more approximate (essentially  
the double $\log$) expression should give a satisfactory estimate of the event rate.  
  
Let us recall the main QCD formulae needed to estimate the inclusive cross section.  The  
partonic quasi-elastic subprocess is $ab \rightarrow a^\prime + {\rm gap} + H + {\rm gap} +  
b^\prime$.  For example, for the subprocess $gg \rightarrow g + H + g$ \cite{KMR}, the  
cross section is given by   
\be    
\label{eq:a5}    
\frac{d \sigma}{dy_{H}} \; = \; A^2 \: \alpha_S^4 \: \frac{81}{2^9 \pi} \: \int     
\frac{dQ^2}{Q^2} \:     
\frac{dQ^{\prime 2}}{Q^{\prime 2}} \: \frac{dt_1}{t_1} \: \frac{dt_2}{t_2} \; e^{- (n_1 +     
n_1^\prime + n_2 + n_2^\prime + S_1 + S_1^\prime + S_2 + S_2^\prime)/2},    
\ee    
where the primes indicate quantities occurring in the complex conjugate amplitude to that    
shown in Fig.~1b.  Now the suppression due to QCD radiative effects comes from the    
double $\log$ resummations $\exp (-n_i/2)$ in the BFKL non-forward amplitudes, as well as    
from the Sudakov form factors $\exp (-S (k_T^2, M_H^2))$ which arise from the    
requirement that there is no gluon emission in the interval $k_T < p_T < M_H/2$.  The    
leading logarithmic contribution again comes from the asymmetric configuration of the  
$t$-channel gluons, $Q_T \ll k_{iT}$.  The amplitude for no gluon emission with $Q_T <    
p_T < k_{iT}$ in the gap $\Delta \eta_i$ is $\exp (-n_i/2)$ where\footnote{There is a second    
$(\ln k_{iT}^2/Q_T^2)$ arising from the BFKL evolution, which when resummed gives the    
BFKL non-forward amplitude $\Phi (\Delta \eta) \exp (-n_i/2)$.  Here $\Delta \eta$ plays the    
role of $\ln (1/x)$ in the BFKL evolution.  For the energies and rapidity gaps of interest this    
BFKL enhancement is small, that is $\Phi (\Delta \eta) \approx 1$.  Of course, a more precise  
calculation to single $\ln Q_T$ accuracy may give a larger amplitude due to less QCD  
radiative suppression.  On the other hand the NLO $\ln (1/x)$ corrections decrease the  
forward $(t_i = 0)$ BFKL amplitude.  Thus we will still take $\Phi (\Delta \eta) \approx 1$.}    
\be    
\label{eq:a6}    
n_i \; = \; \frac{3 \alpha_S}{\pi} \: \Delta \eta_i \: \ln \left ( \frac{k_{iT}^2}{Q_T^2} \right ).    
\ee    
    
The inclusive cross section is obtained by convoluting the parton-parton cross sections with  
the parton densities.  The results are presented in the next section.  
  
\section{Cross sections for exclusive and inclusive Higgs production}  
  
The predictions for the double diffractive Higgs production cross sections are presented in  
Table~1.  The values $\sigma_{\rm excl}$ given for exclusive production are obtained from  
the most complete perturbative QCD calculation that can be made at the present stage.  They  
are based on (\ref{eq:b3}) using the unintegrated gluon distribution of (\ref{eq:g3}).  That is,  
the distributions are calculated to single $\log$ accuracy (as in (\ref{eq:c3}) and  
(\ref{eq:d3})) and include the skewed effect ($R_g$ of (\ref{eq:h3})).  We also include the  
$\alpha_S$ correction to the $gg \rightarrow H$ vertex factor.  That is $A^2$ given by  
(\ref{eq:b2}) is multiplied by the regularized virtual correction \cite{SPIRA}, or so-called  
$K$-factor,  
\be  
\label{eq:z1}  
A^2 \; \rightarrow \; A^2 \: \left ( 1 \: + \: \frac{\alpha_S (M_H^2)}{\pi} \: \left [ \pi^2 \: + \:  
\frac{11}{2} \right ] \right ) \; \simeq \; 1.5~A^2.  
\ee  
  
The predictions of the inclusive cross section $\sigma_{\rm incl}$ in Table~1 are obtained  
using the BFKL non-forward amplitude and are valid to double $\log$ accuracy.  As  
mentioned above, the unintegrated gluon distributions are unknown at large momentum  
transfer $t_i$, and so, at present, we cannot improve our estimates as we have done for the  
exclusive case.  However the values of $\sigma_{\rm incl}$ do include the factor of  
(\ref{eq:z1}).  
  
Finally all the cross sections in Table~1 include the survival probability $S_{\rm spec}^2 =  
0.1$ arising from soft rescattering effects.  As we shall see in the next section, the uncertainty  
in the value of $S_{\rm spec}^2$ gives by far the largest uncertainty in the predicted cross  
sections.  
  
\begin{table}[htb]    
\caption{The cross sections $\left . \sigma = d\sigma/dy_{H} \right |_0$ (in fb) for     
the central production of a Higgs boson in $p\bar{p}$ (or $pp$) collisions at $\sqrt{s} = 2$     
and 14~TeV via the exclusive or inclusive process of Fig.~1 and via $WW$ fusion.  The     
inclusive cross sections are shown for (parton level) rapidity gaps $\Delta \eta = 2$, and also     
for $\Delta \eta = 3$. The tabulated cross sections are obtained using $S_{\rm spec}^2 = 0.1$,  
but see the discussion in Section 5.}    
\begin{center}    
\begin{tabular}{|c|c|c|c|} \hline    
$M_H$ (GeV) & $\sigma_{\rm excl}$ & $\sigma_{\rm incl}$ & $\sigma_{\rm incl} (WW)$     
\\     
& & $\Delta \eta = 2(3)$ & $\Delta \eta = 2(3)$ \\ \hline    
Tevatron $(\sqrt{s} = 2$~TeV) & & & \\    
100 & 0.071 & 1.1 (0.09) & 0.49 (0.031) \\    
120 & 0.030 & 0.62 (0.05) & 0.41 (0.026) \\    
140 & 0.018 & 0.38 (0.03) & 0.35 (0.022) \\    
160 & 0.008 & 0.25 (0.02) & 0.30 (0.019) \\ \hline    
LHC $(\sqrt{s} = 14$~TeV) & & & \\    
100 & 2.4 & 49 (5.5) & 21.6 (5.6) \\    
120 & 1.4 & 36 (3.9) & 20.6 (5.4) \\    
140 & 0.86 & 28 (2.9) & 19.7 (5.2) \\     
160 & 0.55 & 21 (2.3) & 18.8 (5.0) \\ \hline    
\end{tabular}    
\end{center}    
\end{table}  
  
We note that the values of $\sigma_{\rm excl}$ are enhanced in comparison to our previous  
predictions \cite{KMR}, which were based on the low $x$ limit for the unintegrated gluon  
density.  That is we took  
\be  
\label{eq:z2}  
f \; = \; \frac{\partial (xg (x, Q_T^2))}{\partial \ln Q_T^2},  
\ee  
assuming that $\ln (1/x) \gg \ln (M_H/2Q_T)$.  The predictions in Table 1, however, are  
based on the improved expression (\ref{eq:g3}) for the skewed unintegrated gluon density.   
For example at LHC energies both logarithms ($r_x \equiv \ln (1/x) \simeq 4.5$ and $r_T  
\equiv \ln (M_H/2Q_T) \simeq 3.5$) are of comparable size and the derivative of $\sqrt{T}$  
in (\ref{eq:g3}) implies a correction of about $1+ r_T/r_x \simeq 1.8$ to (\ref{eq:z2}), and an  
enhancement of the cross section $d \sigma_{\rm excl} \propto f_g^4$ by a factor $(1.8)^4  
\simeq 11$.  Next we have an enhancement due to the skewed effect, $R_g$ in (\ref{eq:g3}).   
At first sight we might expect that the off-diagonal gluon distribution,  
\be  
\label{eq:z3}  
f_g (x, x^\prime) \; \simeq \; \sqrt{xg (x) x^\prime g (x^\prime)},  
\ee  
would be much larger than the diagonal density, $f_g (x, x)$, since $x^\prime \ll x$ and  
$x^\prime g (x^\prime)$ grows rapidly as $x^\prime \rightarrow 0$.  However it was shown  
\cite{BLOE} that in the leading $\ln (1/x)$ limit  
\be  
\label{eq:z4}  
f_g (x, x^\prime) \; = \; f_g (x, x).  
\ee  
Nevertheless beyond leading $\ln (1/x)$ the ratio is found to be $R_g \simeq 1.2 (1.4)$ at  
LHC (Tevatron) energies, leading to an enhancement $R_g^4 \simeq 2 (4)$, which is  
included in $\sigma_{\rm excl}$ in Table~1.  The third improvement is the inclusion of the  
single logarithmic contribution in (\ref{eq:d3}) for the survival probability $T$.  These  
contributions allow for the kinematic constraints on gluon emission and enlarge the value of  
$T$.  (It is well known that the double $\log$ expression of the type of that was used in  
\cite{KMR}, overestimates the suppression.)  
  
Table 2 shows how the QCD prediction for the exclusive cross section $\sigma_{\rm excl}$  
would be reduced if we omit the various improvements one-by-one.  $\sigma_{\rm excl}$  
becomes $\sigma_1$ if we switch off the single $\log$ contribution to $T$ in (\ref{eq:d3})  
and return to the double $\log$ formula of \cite{KMR}.  If we then omit the $\partial  
T/\partial \ln Q_T^2$ term in (\ref{eq:c3}, \ref{eq:g3}) for $f_g$ the cross section reduces to  
$\sigma_2$, and finally if we omit the skewed effect factor $R_g$ we obtain $\sigma_3$.  
  
\begin{table}[htb]    
\caption{The reduction in the cross section $\sigma_{\rm excl} \rightarrow \sigma_1  
\rightarrow \sigma_2 \rightarrow \sigma_3$ due to the omission of the various QCD  
improvements one-by-one, as detailed in the text.  $\left . \sigma \equiv d \sigma/dy_H \right  
|_0$ in fb.}    
\begin{center}    
\begin{tabular}{|c|cc|cc|} \hline  
& \multicolumn{2}{|c|}{Tevatron ($\sqrt{s} = 2$~TeV)}   
& \multicolumn{2}{|c|}{LHC ($\sqrt{s} = 14$~TeV)} \\   
$M_H$ (GeV) & 120 & 160 & 120 & 160 \\ \hline  
$\sigma_{\rm excl}$ & 0.030 & 0.008 & 1.4 & 0.55 \\   
$\sigma_1$ & 0.012 & 0.003 & 0.56 & 0.21 \\  
$\sigma_2$ & $0.8 \times 10^{-4}$ & $0.8 \times 10^{-5}$ & 0.03 & 0.007 \\  
$\sigma_3$ & $0.2 \times 10^{-4}$ & $0.2 \times 10^{-5}$ & 0.012 & 0.003 \\ \hline  
\end{tabular}    
\end{center}    
\end{table}  
  
\indent We should discuss the $Q_T^2$ integration which is necessary to compute the  
exclusive amplitude (\ref{eq:b3}).  We take the lower limit to be $Q_0^2 = 1.25$~GeV$^2$,  
since this is the starting value of the MRS \cite{MRSR} partons that we use.  The saddle  
points of the $d \ln Q_T^2$ integration are at about $Q_{\rm SP}^2 = 3.2$~GeV$^2$ and  
1.5~GeV$^2$ for the LHC and Tevatron energies respectively.  
  
The prediction for the rapidity distribution for double diffractive exclusive Higgs production  
is shown in Fig.~2.  After integration over rapidity we obtain\footnote{This value may be  
compared to the original estimate implied by Bialas and Landshoff \cite{BL} of  
$\sigma_{\rm excl} \simeq 100$~fb, which should be multiplied by $8 S_{\rm spec}^2  
\simeq 1$.  The factor of 8 is necessary to allow for the identity of the gluons.} $\sigma_{\rm  
excl} = 5.7$~fb for $M_H = 120$~GeV at $\sqrt{s} = 14$~TeV.  
  
We see from Table~1 that the values of the cross section for the inclusive process   
(\ref{eq:a4}) are much larger than the exclusive cross section.  Recall that the reasons are that  
(i) the QCD suppression is not so strong and (ii) the Higgs boson may be produced with a  
larger transverse momentum.  However we see that for a large rapidity gap, $\Delta \eta = 3$,  
the inclusive rate falls to a value comparable to that for the exclusive process.  
  
Up to now we have dealt with a purely perturbative QCD expression.  All the next-to-leading  
kinematically enhanced effects are under control and so we may hope that the higher order  
$\alpha_S$ effects will only give a 20--40\% correction. Much more uncertainty comes from  
the soft non-perturbative effects.  First there is the contribution coming from $Q_T^2 <  
Q_0^2 = 1.25$~GeV$^2$.  We can estimate it using GRV partons \cite{GRV} (where we  
may take $Q_0 \simeq 0.6$~GeV) or by freezing the anomalous dimension of the MRS gluon  
for $Q_T < Q_0$, but still allowing $\alpha_S$ to run in (\ref{eq:d3}).  In both cases the  
contributions are comparable and not too large at the LHC energy; $\sigma_{\rm excl}$  
increases by $\lapproxeq 20$~\% for $M_H = 120-160$~GeV. Note that it is the inclusion of  
the Sudakov suppression effects which makes the perturbative QCD estimates infrared stable  
and hence the predictions reliable.  On the other hand at the Tevatron energy the position of  
the saddle point is rather low ($Q_{\rm SP}^2 \simeq 1.5$~GeV$^2$) and the contribution  
from $Q_T < Q_0$ enlarges the cross section by about a factor of 2.  
  
The main uncertainty, however, does not come from any of the above effects, but arises from  
the survival probability of the rapidity gaps with respect to the soft rescattering effects.  As  
we shall see below it appears that we have been too optimistic to use in Table~1 the canonical  
value $S_{\rm spec}^2 = 0.1$ at LHC energies.  
  
\section{Suppression due to soft rescattering}  
  
Here we will show that the main uncertainty in the calculation of the double diffractive cross  
sections arises from the estimate of how much soft rescattering fills up the rapidity gaps; that  
is, in the estimate of the probability, $S_{\rm spec}^2$, for soft rescattering {\it not} to  
occur.  
  
To begin, we consider the effect of a single elastic rescattering, which is shown in Fig.~3a  
where the blob represents the whole $pp$ (or $p\bar{p}$) elastic amplitude including the  
absorptive corrections.  It is straightforward to show that this simple rescattering amplitude  
gives a survival probability  
\be  
\label{eq:s0}  
S_{\rm spec} \; = \; \left ( 1 \: - \: \frac{\sigma_{\rm tot}}{4 \pi (B_{\rm el} + 2b)} \right )  
\ee  
where $B_{\rm el}$ is the slope of the elastic differential cross section $(d\sigma/dt \sim \exp  
(B_{\rm el} t))$, $b$ determines the $t$ dependence of exclusive Higgs production (via the  
proton form factors $\sim \exp (bt_i/2)$ in the amplitude) and $\sigma_{\rm tot}$ is the $pp$  
(or $p\bar{p}$) total cross section.  For example at LHC energies, where we expect  
$\sigma_{\rm tot} \approx 100$~mb and $B_{\rm el} \approx 20$~GeV$^{-2}$,  
(\ref{eq:s0}) gives  
\be  
\label{eq:s0a}  
S_{\rm spec} \; \approx \; 0.65,  
\ee  
if we take $b = 5.5$~GeV$^{-2}$ as before.  
  
To allow for dissociation in the rescattering process (shown by the heavier intermediate states  
in Fig.~3b) we multiply $\sigma_{\rm tot}$ by a factor $C^2 > 1$, where  
\be  
\label{eq:sA}  
C \; = \; 1 \: + \: \frac{\sigma ({\rm target~dissociation})}{\sigma _{\rm el}} \; .  
\ee  
Here $\sigma_{\rm el}$ is the elastic $pp$ cross section.  Note that, for the above example, if  
$C$ becomes greater than 1.25, then we would have already obtained a negative value for  
$S_{\rm spec}$.  This is a warning that we need to be careful in the precise values that we  
assume for $\sigma_{\rm tot}, B_{\rm el}$ and $C$ at LHC energies.  Alternatively, we can  
sum up the effect of multiple rescattering using a model which embodies unitarity and  
therefore has $S_{\rm spec} > 0$ built in.  
  
Let us consider the eikonal model sketched in Fig.~3c.  In this model we have  
\bea  
\label{eq:s1}  
\sigma_{\rm tot} & = & \frac{2}{C^2} \: \int \: d^2 \rho_T \: \left (1 - e^{- \Omega  
(\rho_T)/2} \right ), \\  
& & \nonumber \\  
\label{eq:s2}  
\sigma_{\rm el} & = & \frac{1}{C^4} \: \int \: d^2 \rho_T \: \left (1 - e^{- \Omega  
(\rho_T)/2} \right )^2  
\eea  
where the impact parameter $\rho_T$ is the transverse coordinate of the incoming proton  
with respect to the target proton, and $\Omega (\rho_T)$ is the optical density (or opacity) of  
the interaction.  Here $\exp (- \Omega)$ reflects the absorption of the incoming beam, and  
$\exp (- \Omega/2)$ describes the reduction of the amplitude at impact parameter $\rho_T$.   
Thus  
\be  
\label{eq:s2a}  
S_{\rm spec} \; = \; \langle e^{- \Omega (\rho_T)/2} \rangle  
\ee  
where the average is taken over the $\rho_T$ dependence, $\exp (- \rho_T^2/4b)$, of the  
amplitude for exclusive Higgs production  
\be  
\label{eq:sB}  
{\cal M} \; \propto \; e^{b (t_1 + t_2)/2}.  
\ee  
That is we have  
\be  
\label{eq:s3}  
S_{\rm spec} \; = \; \frac{\int \: d^2 \rho_T \: e^{- \Omega (\rho_T)/2} \: e^{-  
\rho_T^2/4b}}{\int \: d^2 \rho_T \: e^{- \rho_T^2/4b}}.  
\ee  
For the opacity we take the Gaussian form  
\be  
\label{eq:s4}  
\Omega (\rho_T) \; = \; \frac{C^2 \sigma (s)}{2 \pi B} \: e^{- \rho_T^2/2B},  
\ee  
where $\sigma (s) = \sigma_0 (s/s_0)^\Delta$ corresponds to the Pomeron exchange  
amplitude shown by the double lines in Fig.~3c.  We take the slope of the Pomeron amplitude  
to be  
\be  
\label{eq:s5}  
\frac{B}{2} \; = \; B_0 \: + \: 2 \alpha_P^\prime \ln (s/s_0).  
\ee  
We tune the parameters $(\sigma_0, B_0, \Delta, \alpha_P^\prime)$ of the eikonal model to  
describe the behaviour of $\sigma_{\rm tot}$ and the $pp$ elastic scattering data throughout  
the ISR to Tevatron energy range ($30 < \sqrt{s} < 1800$~GeV).  Finally we study the  
predictions for $S_{\rm spec}$ for two relevant values of the enhancement parameter $C$ of  
(\ref{eq:sA}), namely $C = 1.15$ and $C = 1.3$.  The smaller value of $C$ is obtained if we  
include only the nucleon resonance excitations \cite{KAID}; in terms of partons it means that  
we account mainly for valence quark rescattering.  On the other hand at the larger (LHC)  
energies we have to include rescattering of partons with small $x$ (\lq wee\rq~partons), and  
in this case $C \approx 1.3$ is more appropriate.  Both choices allow a satisfactory fit of  
$\sigma_{\rm tot}$ and the elastic data\footnote{The parameter $\Delta$, which specifies the  
Pomeron intercept, is found to be $\Delta = 0.10$ and $\Delta = 0.13$ in order to fit the \lq\lq  
soft\rq\rq~data, taking $C = 1.15$ and $C = 1.3$ respectively.}, and the values of $S_{\rm  
spec}^2$ obtained from (\ref{eq:s3}) are shown in Table 3.  
  
\begin{table}[htb]    
\caption{The probability $S_{\rm spec}^2$, that the rapidity gaps survive rescattering,  
calculated using the eikonal model (\ref{eq:s3}) for two values of the enhancement factor  
$C$ of (\ref{eq:sA}), namely $C = 1.15$ and $C = 1.3$ (expected to be appropriate for  
Tevatron and LHC energies respectively).  The slope $b$ for exclusive Higgs production,  
(\ref{eq:sB}), is expected to be 5.5~GeV$^{-2}$, but smaller values are not excluded (see  
text).}    
\begin{center}    
\begin{tabular}{|c|cc|cc|} \hline    
& \multicolumn{2}{|c|}{Tevatron $(\sqrt{s} = 2$~TeV)}  
& \multicolumn{2}{|c|}{LHC $(\sqrt{s} = 14$~TeV)}\\     
\cline{2-5} \raisebox{1.5ex}[0pt]{$b$ GeV$^{-2}$} & $C = 1.15$ &  $C = 1.3$ & $C =  
1.15$ & $C = 1.3$ \\ \hline    
5.5 & 0.11 & 0.034 & 0.054 & 0.011 \\   
4 & 0.07 & 0.013 & 0.029 & 0.003 \\   
2.5 & 0.04 & 0.003 & 0.012 & 0.0003 \\ \hline    
\end{tabular}    
\end{center}    
\end{table}  
  
\indent From the results shown in Table~3 we see that the survival probability $S_{\rm  
spec}^2$ depends sensitively on the value of the slope $b$ (of (\ref{eq:sB})), that is on the  
spatial $(\rho_T)$ distributions of partons inside the proton, see (\ref{eq:s3}).  Unfortunately  
we cannot be completely sure that the value we have adopted, $b = 5.5$~GeV$^{-2}$, is  
correct.  This value is obtained by assuming that the distribution of colour dipoles (gluons)  
mimicks the electric charge distribution of the proton.  However in diffractive $J/\psi$  
electroproduction the slope is observed to be $b_\psi \approx 4$~GeV$^{-2}$ \cite{HERA}.   
Since this process is also mediated by two-gluon exchange, another choice for $b$ could be  
4~GeV$^{-2}$.  Moreover the $J/\psi$ slope is given by  
\be  
\label{eq:s6}  
b_\psi \; = \; b_0 \: + \: 2 \alpha_P^\prime \ln (1/x_\psi),  
\ee  
where $1/x_\psi \approx W^2/M_\psi^2$.  Thus, since the $J/\psi$ HERA data \cite{HERA}  
sample $x \sim 10^{-3}$, whereas Higgs production at the LHC samples $x \sim 10^{-2}$, it  
suggests that $b$ is about ($2 \alpha_P^\prime \ln 10$) less than $b_\psi$.  For this reason we  
also show results in Table~3 for $b = 2.5$~GeV$^{-2}$, which if it were true would give  
much lower values of $S_{\rm spec}^2$.  However the evidence from HERA is conflicting.   
Large $Q^2$ open $q\bar{q}$ or $\rho$ meson electroproduction \cite{HERA} are observed  
to have slope $b_\rho \approx 7$~GeV$^{-2}$, in the same region of $x$ as $J/\psi$.  From  
this point of view the lower $J/\psi$-motivated values of $b$ look anomalous.  So, on  
balance, the value $b = 5.5$~GeV$^{-2}$ looks to be the most realistic.  
  
Taking $b = 5.5$~GeV$^{-2}$, we see from Table~3, that at the Tevatron we have $S_{\rm  
spec}^2 \approx 0.1$, which is the canonical value that we have used.  However at the  
LHC\footnote{Note that the simple formula (\ref{eq:c2}) would give $S_{\rm spec}$  
independent of collider energy if $\sigma^D/\sigma_{\rm tot}$ were constant, which is not  
incompatible with the present data for $\sigma^D/\sigma_{\rm tot}$, although the errors are  
large.  However this estimate is too naive since it assumes the same spatial distribution of  
partons in the proton for both the soft and hard processes, that is $b = B_{\rm el}/2$.} the  
corresponding value is $S_{\rm spec}^2  \approx 0.05$.  Even worse, for the more relevant  
choice of enhancement factor, $C = 1.3$ at LHC energies, we predict $S_{\rm spec}^2  
\approx 0.01$.  
  
We have used other models to estimate the survival probability $S_{\rm spec}^2$ and, given  
the values of $C$ and $b$, we have found essentially the same results as in Table~3.  The  
reason is interesting.  At high energies the centre of the proton becomes black, that is  
$\Omega \gg 1$ and $\exp (- \Omega/2) \simeq 0$ in (\ref{eq:s3}).  Hence the main  
contribution to $S_{\rm spec}$ comes from the peripheral region $\rho_T \gapproxeq  
\rho_0$, where $\rho_0$ is where the proton starts to become transparent, i.e. $\Omega  
(\rho_T \gapproxeq \rho_0) \lapproxeq 1$.  Thus, as long as the model describes  
$\sigma_{\rm tot}$ and elastic $pp$ data, the prediction for $S_{\rm spec}$ does not depend  
crucially on the details, but is controlled essentially by the proton radius (or $B_{\rm el}  
\approx \sigma_{\rm tot}^2/16 \pi \sigma_{\rm el})$ and the slope $b$.  
  
Let us finally comment on $S_{\rm spec}^2$ for a Higgs boson produced by $WW$ or  
$\gamma\gamma$ fusion.  It is not excluded that the radius of the quark distributions in the  
proton is larger than that of the gluons.  If this were the case, then $S_{\rm spec}^2$ for  
$WW$ fusion would be larger than those shown in Table~3.  The most exciting example is  
$\gamma\gamma \rightarrow H$ production.  This process takes place at very large impact  
parameter $\rho_T$, and here $S_{\rm spec} \simeq 1$.  In Ref.~\cite{GAMMA} the  
$\gamma\gamma \rightarrow H$ cross section at the LHC was predicted to be 0.3~fb for  
$M_H \approx 120$~GeV.  This would be close to our prediction of 0.6~fb for production by  
two-gluon exchange if we were to take a survival factor of $S_{\rm spec}^2 = 0.01$, instead  
of $S_{\rm spec}^2 = 0.1$.  If indeed this is the case and, moreover, if we were to assume  
that $b < 5.5$~GeV$^{-2}$ is correct, then we would have the following hierarchy  
\be  
\label{eq:s7}  
1 \; \approx \; S_{\rm spec}^2 (\gamma\gamma \rightarrow H) \; \gg \; S_{\rm spec}^2 (WW  
\rightarrow H) \; > \; S_{\rm spec}^2 (\funp\!\funp \rightarrow H),  
\ee  
where $\funp\!\funp$ denotes the two-gluon exchange mechanism of Fig.~1.  Of course for 
the default choice $b = 5.5$~GeV$^{-2}$ (i.e.\ assuming the spatial distributions of quarks 
and gluons to be the same) we have 
\be 
\label{eq:s8} 
\; S_{\rm spec}^2 (WW \rightarrow H) \; = \; S_{\rm spec}^2 (\funp\!\funp \rightarrow H). 
\ee 
  
\section{The dijet monitor}  
    
The previous section demonstrates that the \lq\lq Achilles heel\rq\rq~of the calculation of the  
Higgs production cross sections is the uncertainty in the {\it soft} survival factor $S_{\rm  
spec}^2$.  Fortunately there is a way to experimentally measure $S_{\rm spec}^2$ by  
observing the double diffractive production of a pair of high $E_T (\sim M_H/2)$ jets with  
rapidity gaps on either side of the pair.  The process is described by the same Feynman  
diagrams, both in the case of the exclusive process (\ref{eq:a1}) and also for inclusive  
production (\ref{eq:a4}).  Essentially we need simply to replace the $gg \rightarrow H$  
subprocess by that for $gg \rightarrow$ dijet.  The dijet event rate is much larger than that for  
the Higgs signal and so the collider experiments should be able to directly test the QCD  
estimates and measure $S_{\rm spec}^2$.    
    
QCD estimates of the rapidity gap dijet rate were given in \cite{MRK}.  In Table~4 we     
present improved numerical results in a kinematic range comparable to that for Higgs  
production.  We use the same prescription that was employed to calculate the Higgs  
production cross sections presented in Table~1.  That is for exclusive dijet production we  
integrate from $Q_T = Q_0$ to $Q_T = E_T$ over skewed unintegrated gluons (\ref{eq:g3}),  
calculating the QCD radiative survival factor $T$ to single $\log$ accuracy.  However the  
NLO $K$-factor for the subprocess $gg \rightarrow\:$dijet is omitted.  This correction  
depends on the \lq\lq jet finding\rq\rq~algorithm.  Usually the size of the jet cone is chosen in  
such a way that the effective NLO $K$-factor is close to 1.  For comparison with Table~1 we  
continue to use the canonical soft rescattering factor $S_{\rm spec}^2 = 0.1$, although we  
note from Section 5 the true factor may be smaller.  
  
In practice it is impossible to study a purely exclusive dijet production process, analogous to  
(\ref{eq:a1}).  We cannot distinguish a bremsstrahlung gluon emitted in the dijet rapidity  
interval from a gluon which belongs to one of the high-$E_T$ jets.  We have therefore chosen  
rapidity gaps such that bremsstrahlung is only forbidden for $| \eta_g | > 2$ in the dijet centre- 
of-mass frame.  Of course the QCD radiative suppression will be much smaller in this case.   
For example, for semi-exclusive production of $E_T = 50$~GeV jets at LHC energies, we  
have a typical survival factor $T (Q_{\rm SP}, M_H/2) \simeq 0.5$ at the saddle point  
$Q_{\rm SP}^2 \simeq 2$~GeV$^2$ of the $d \ln Q_T^2$ integration.  
  
The double diffractive dijet cross sections are much larger than those for Higgs production.   
For example if we take a dijet bin of size $\delta E_T = 10$~GeV for each jet and $\delta  
(\eta_1 - \eta_2) = 1$ we estimate, for $E_T = 50$~GeV jets at LHC energies,  
\be  
\label{eq:z5}  
\left . d \sigma_{\rm excl}/d \eta \right |_0 \; \simeq \; 38~{\rm pb}, \quad\quad \left . d  
\sigma_{\rm incl}/d\eta \right |_0 \; \simeq \; 240~{\rm pb}  
\ee  
where $\eta \equiv (\eta_1 + \eta_2)/2$, and the rapidity gaps are taken to be $\Delta \eta ({\rm  
veto}) = (\eta_{\rm min}, \eta_{\rm max}) = (2, 4.1)$ for the inclusive case (see \cite{MRK}  
for the definition of the dijet kinematics).  For 30~GeV jets at the Tevatron the corresponding  
cross sections are  
\be  
\label{eq:z6}  
\left . d \sigma_{\rm excl}/d \eta \right |_0 \; \simeq \; 17~{\rm pb}, \quad\quad \left . d  
\sigma_{\rm incl}/d \eta \right |_0 \; \simeq \; 150~{\rm pb}.  
\ee  
  
The numbers in square brackets in Table~4 correspond to double diffractive $b\bar{b}$ dijet  
production.  For $M_H$ in the range that we consider, this process is the main background to  
the double diffractive Higgs signal.  However the event rate of $b\bar{b}$ jets is more than  
two orders of magnitude lower than the gluon dijet rate.  Even after integration over a $\delta  
E_T = 10$~GeV interval the rate is comparable to the Higgs cross section.  We conclude the  
$b\bar{b}$ background should not be a problem.  
  
We emphasize that the rapidity intervals chosen in our calculations refer to partonic  
rapidities.  Table 4 shows that the cross sections depend sensitively on the size of the rapidity  
gaps, $\Delta \eta_{\rm veto}$.  Practical estimates require Monte Carlo simulations  
appropriate to the specific experimental cuts and which include treatment of possible initial  
state radiation and the hadronization of the large $E_T$ jets.  
    
\begin{table}[htb]    
\caption{The cross sections $\left . \sigma = d\sigma/dp_T^2 d\eta d \Delta \eta\right |_{\eta     
= 0}$ (in fb/GeV$^2$) for the exclusive and inclusive double diffractive dijet production, for  
three values of the rapidity difference of the two jets $\Delta \eta \equiv \eta_1 - \eta_2$. The  
results are shown for different collider energies $(\sqrt{s})$ and different transverse energy  
$(E_T)$ of the jets.  The two jets are taken to have the same $E_T$.  The rapidity gaps are     
defined by the intervals $\pm \Delta \eta_{\rm veto} \equiv \pm (\eta_{\rm min}, \eta_{\rm  
max})$ \cite{MRK}.  The numbers in the square brackets are the $b\bar{b}$ component of  
the dijet signal.}    
\begin{center}    
\begin{tabular}{|c|c|c|c|c|} \hline    
& & $\sigma_{\rm excl} (jj)$ & \multicolumn{2}{|c|}{$\sigma_{\rm incl} (jj) \quad  
[\sigma_{\rm incl} (b\bar{b})]$} \\ \cline{4-5}   
& \raisebox{1.5ex}[0pt]{$\Delta \eta$} & $[\sigma_{\rm excl} (b\bar{b})]$ &  $\Delta  
\eta_{\rm veto} = (2, 4.1)$ & $\Delta \eta_{\rm veto} = (1.5, 4.6)$ \\ \hline    
$\sqrt{s} = 2$~TeV & 0 & 0.97 [0.008] & 5.2 [0.04] & 0.28 [0.002] \\   
$E_T = 50$~GeV & 1 & 0.76 [0.005] & 4.4 [0.03] & 0.23 [0.0015] \\  
& 2 & 0.31 [0.001] & 2.4 [0.01] & 0.11 [0.0004] \\ \hline    
$\sqrt{s} = 2$~TeV & 0 & 29 [0.23] & 240 [1.9] & 15 [0.12] \\  
$E_T  = 30$~GeV & 1 & 22 [0.13] & 220 [1.5] & 13 [0.09] \\  
& 2 & 11 [0.03] & 140 [0.5] & 8 [0.3] \\ \hline  
$\sqrt{s} = 14$~TeV & 0 & 38 [0.30] & 240 [1.9] & 24 [0.19] \\  
$E_T = 50$~GeV & 1 & 31 [0.22] & 240 [1.6] & 23 [0.16] \\  
& 2 & 19 [0.08] & 200 [0.7] & 18 [0.07] \\ \hline  
\end{tabular}    
\end{center}    
\end{table}  
  
\section{Comparison with other QCD-based predictions}  
  
We have argued that perturbative QCD gives reliable estimates for double diffractive Higgs   
and dijet production, up to the uncertainty in the soft rescattering effects.  It is therefore  
important to understand the origin of the difference with recent more optimistic estimates of  
the event rates.  
  
Double diffractive high $E_T$ dijet production has been recently estimated by Berera  
\cite{B}.  His non-factorized N(L)DPE model is similar to our perturbative QCD approach.    
However there are some differences in application.  First, in \cite{B} a fixed coupling   
$\alpha_S (E_T^2)$ was used in the double $\log$ form of the Sudakov suppression factor,  
whereas here (and in Ref.~\cite{MRK}) we use the more appropriate running coupling  
$\alpha_S (p_T^2)$ inside the integration $Q_T^2 < p_T^2 < E_T^2$.  Second, dijet  
production cannot be a pure exclusive process since it is impossible to forbid extra emission  
in the central rapidity interval occupied by the dijets.  Thus we have, at best, a semi-exclusive  
reaction in which the suppression\footnote{So, the Sudakov suppression appears to be much  
weaker for dijet production than for the analogous Higgs process.  We can easily gain insight  
into the origin of this difference by recalling a similar phenomenon in the radiative effects  
accompanying the production of narrow and wide heavy resonances (see, for example, the  
discussion of well known QED effects in Ref.~\cite{QED}).  The energetic bremsstrahlung  
pushes the initial state off the resonance energy for the non-radiative process.  Thus, the  
narrow resonance could be produced only if we damp radiation with energy exceeding the  
resonance width.  The wider the resonance, the larger is the phase space available for  
emission and, therefore, the less pronounced is the Sudakov suppression.} is only relevant in  
some rapidity interval $\delta \eta_{\rm veto}$, see Ref.~\cite{MRK}.  When this is taken  
into account the predictions \cite{B} are reduced by an amount that roughly compensates for  
the enhancement due to the use of $\alpha_S (E_T^2)$.  The most important numerical  
difference between \cite{B} and our predictions is due to treatment of the gluon exchanges.   
In Ref.~\cite{B} (and also in \cite{EML}, as we discuss below) a non-perturbative two-gluon  
model is used in which the gluon propagator is modified so as to reproduce the total cross  
section.  On the contrary, we have used a realistic unintegrated gluon density, determined  
from conventional gluons of global parton analyses, which has been found to give a  
consistent description of other processes described by perturbative QCD.  
  
It was emphasized in \cite{B} (see also \cite{MRK}) that the above non-perturbative  
normalisation based on the value of the elastic (or total) cross section fixes the diagonal gluon  
density at $\hat{x} \sim \ell_T/\sqrt{s}$ where the transverse momentum $\ell_T$ is small,  
namely $\ell_T < 1$~GeV~$< Q_0$.  Thus the value of $\hat{x}$ is even smaller than  
\be  
\label{eq:a7}  
x^\prime \; \approx \; Q_T/\sqrt{s} \; \ll \; x \; \approx \; M_H/\sqrt{s}.  
\ee  
However, the gluon density grows as $x \rightarrow 0$ and so it is clear that such a non- 
perturbative gluon normalisation will overestimate the double diffractive cross section.  
  
We now turn to the very optimistic (\lq\lq brave\rq\rq) estimate for double diffractive  
exclusive Higgs production, process (\ref{eq:a1}), that has recently been presented in  
\cite{EML}.  For $M_H = 100$~GeV the prediction is $d \sigma/dy \simeq 20$~fb, even for  
Tevatron energies, see Fig.~2.  We are unable to justify this estimate.  First the radiative  
suppression $T^2$ ($\equiv S_{\rm par}^2$ in \cite{EML}) $\simeq 0.1$ which is much  
larger than our determination.  The phenomenological estimate of $T^2$ in \cite{EML} is  
based on the known hadron multiplicity $N_{\rm had}$ measured in the rapidity interval  
$\Delta y = \ln (M_H^2/s_0)$ in a {\it soft} hadron-hadron collision.  This multiplicity  
increases as $\ln (M_H^2/s_0)$ and has nothing to do with the double logarithmic  
bremsstrahlung, where the mean number of emitted gluons $S \propto \ln^2 M_H^2$.  Next,  
motivated by the BLM prescription \cite{BLM}, the coupling $\alpha_S$ in the $gg  
\rightarrow H$ matrix element (that is in the analogue of (\ref{eq:b2})) is evaluated in  
\cite{EML} at a low scale $Q_0$ rather than $M_H$.  However to apply the BLM procedure  
consistently we must ascertain which part of the gluon   
self-energy insertions are already included in the survival factor $T$ (or $S_{\rm par}^2$),  
and which part should be attributed to $\alpha_S$.  Indeed in calculating $T$ to single $\log$  
accuracy we obtained the pre-exponential factor  
\be  
\label{eq:a8}  
\left [ \frac{\ln (Q_T^2/\Lambda_{\rm QCD}^2)}{\ln (M_H^2/4 \Lambda_{\rm QCD}^2)}  
\right ]^{-1} \; = \; \frac{\alpha_S (Q_T^2)}{\alpha_S (M_H^2/4)},  
\ee  
see (\ref{eq:f3}).  This factor reflects the fact that, as usual, the double $\log$ approximation  
overestimates the kinematically available phase space for emission.  That is the probability  
not to bremsstrahlung a gluon, $T$, is larger than $\exp (-S)$.  Thus, in conclusion, as far as  
the single $\log$ corrections are already included in the $T$ factor, we must use $\alpha_S  
(M_H^2)$ in (\ref{eq:b2}), together with the $K$-factor of (\ref{eq:z1}) evaluated at scale  
$\mu = M_H$.  
  
From Table~1 and Fig.~2 we see that the perturbative QCD predictions for $d\sigma/dy_H$  
show a strong increase with increasing energy, which arises because of the growth of the  
gluon densities $xg (x, Q_T^2)$ with increasing $1/x \simeq s/M_H^2$.  On the contrary the  
non-perturbative two-gluon-exchange-type phenomenological models have no $x$  
dependence.  The predictions of these models depend only weakly on energy through the  
energy dependence of the \lq\lq soft\rq\rq~cross section which is used to normalise the   
two-gluon exchange contribution.  The same arguments apply to the production of a pair of  
high $E_T$ jets.  Therefore an experimental study of the dijet production rate as a function of  
the collider energy will clearly be able to discriminate between the perturbative QCD  
determinations and the non-perturbative model approaches.  
  
\section{Summary}  
  
We have calculated the cross sections for exclusive and inclusive double diffractive Higgs  
boson, and also dijet, production in the central region, at both LHC and Tevatron energies.   
That is  
\bea  
\label{eq:y1}  
pp & \rightarrow & p \; + \; (H \; {\rm or} \; jj) \; + \; p \nonumber \\  
& & \\  
pp & \rightarrow & X \; + \; (H \; {\rm or} \; jj) \; + \; Y \nonumber  
\eea  
where $+$ denotes a rapidity gap.  These processes are driven by \lq asymmetric\rq~two  
gluon exchange, with the colour screening gluon being comparatively soft, but still in the  
perturbative QCD domain.  
  
All the important perturbative QCD corrections were included in the calculation.  A  
prescription for the unintegrated gluon distribution, up to single log accuracy, was used.  The  
major uncertainty comes from the non-perturbative sector, namely from the value of the  
survival factor $S_{\rm spec}^2$ --- the small probability to have no secondaries from soft  
rescattering populating the rapidity gaps.  We found that this probability $S_{\rm spec}^2$  
depends sensitively on the spatial distribution of gluons inside the proton.  In the tables in  
which we presented the Higgs and dijet cross section predictions, we took $S_{\rm spec}^2 =  
0.1$, but at LHC energies our estimates of $S_{\rm spec}^2$ indicate that the value is most  
likely to be an order of magnitude smaller.    
  
To overcome the normalisation uncertainty due to the lack of knowledge of $S_{\rm  
spec}^2$ we proposed that measurements of the double diffractive production of dijets would  
act as a luminosity monitor for the two-gluon exchange processes of (\ref{eq:y1}).  The  
estimates of the dijet cross section are such that the process should be readily observable at  
the Tevatron and at the LHC.  In particular measurements of jets with $E_T \sim M_H/2$  
would enable the cross section for the double diffractive production of the Higgs boson to be  
reliably predicted, since the two processes are driven by the same $S_{\rm spec}$ factor in  
the same kinematic region.  Unfortunately even our most optimistic predictions for the Higgs  
process are considerably smaller than previous estimates, and would make the process hard to  
observe at the Tevatron and the LHC.  
  
\section*{Acknowledgements}  
  
We thank M. Albrow, A. Berera, J. Ellis, R. Hirosky and E. Levin for useful discussions.   
VAK thanks the Leverhulme Trust for a Fellowship.  This work was also supported by the  
Royal Society, PPARC, the Russian Fund for Fundamental Research (98-02-17629) and the  
EU Framework TMR programme, contract FMRX-CT98-0194 (DG 12-MIHT).

\newpage  
\noindent {\large \bf Figure Captions}  
  
\begin{itemize}  
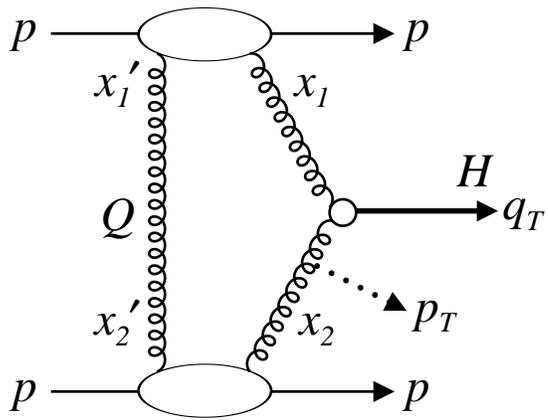
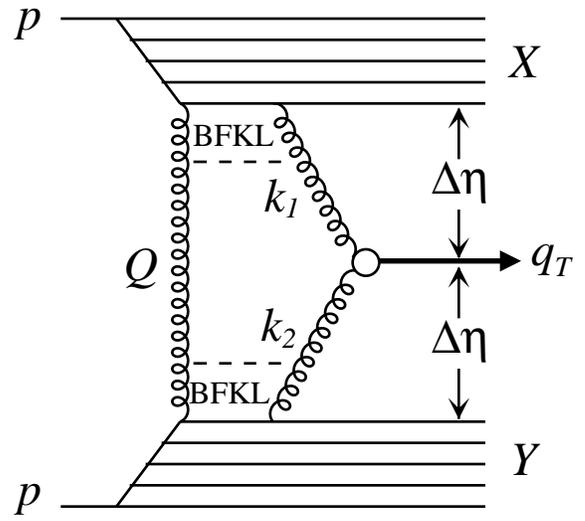
\item[Fig.~1] Diagrams for (a) exclusive, and (b) inclusive, double diffractive Higgs   
production of transverse momentum $q_T$ in high energy $pp$ (or $\bar{p}p$) collisions.    
The QCD radiative corrections (such as the emission of the gluon of transverse momentum   
$p_T$) suppress the number of rapidity gap events via Sudakov form factors, $\exp (-S)$.   
For   
inclusive production $q_T$ can be much larger and the Sudakov suppression is weaker;  
however there are additional QCD radiative effects from the double $\log$ resummations   
$\exp (-n_i/2)$ in the BFKL non-forward amplitudes.  
  
\item[Fig.~2] The perturbative QCD predictions of the rapidity distribution for double  
diffractive exclusive Higgs production at LHC and Tevatron energies.  We also show the  
recent prediction by Levin \cite{EML}, which is discussed in Section 7.  
  
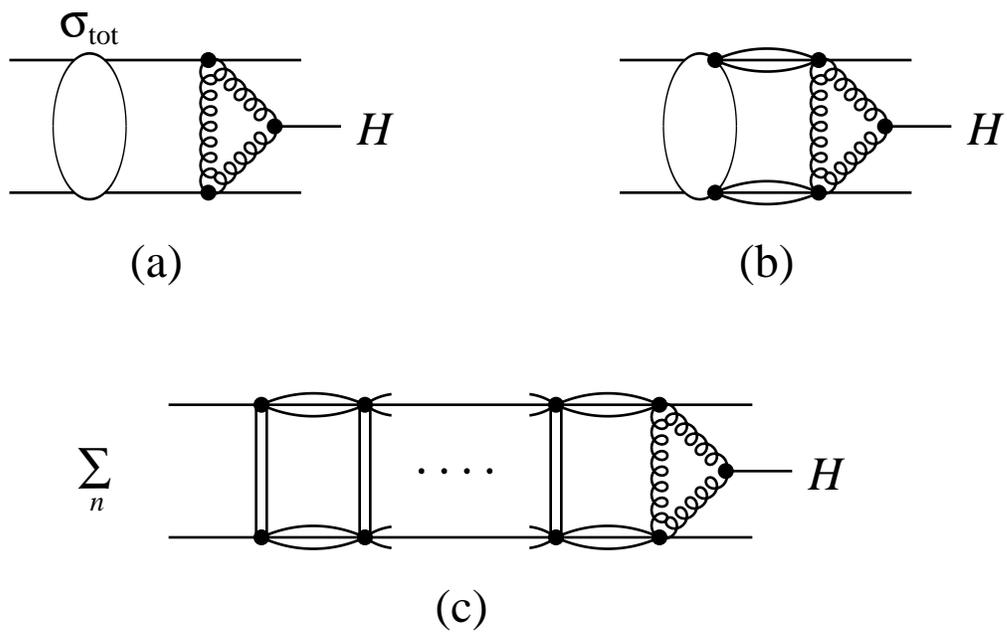
\item[Fig.~3] Diagram (a) illustrates the absorptive correction to exclusive Higgs production,  
assuming that only elastic $pp$ rescattering occurs, with an amplitude Im$A = s \sigma_{\rm  
tot}$.  Diagram (b) includes both elastic and inelastic intermediate states.  Diagram (c) is an  
eikonal representation of (b), where the double line denotes Pomeron exchange.  
  
\end{itemize}   
                                    
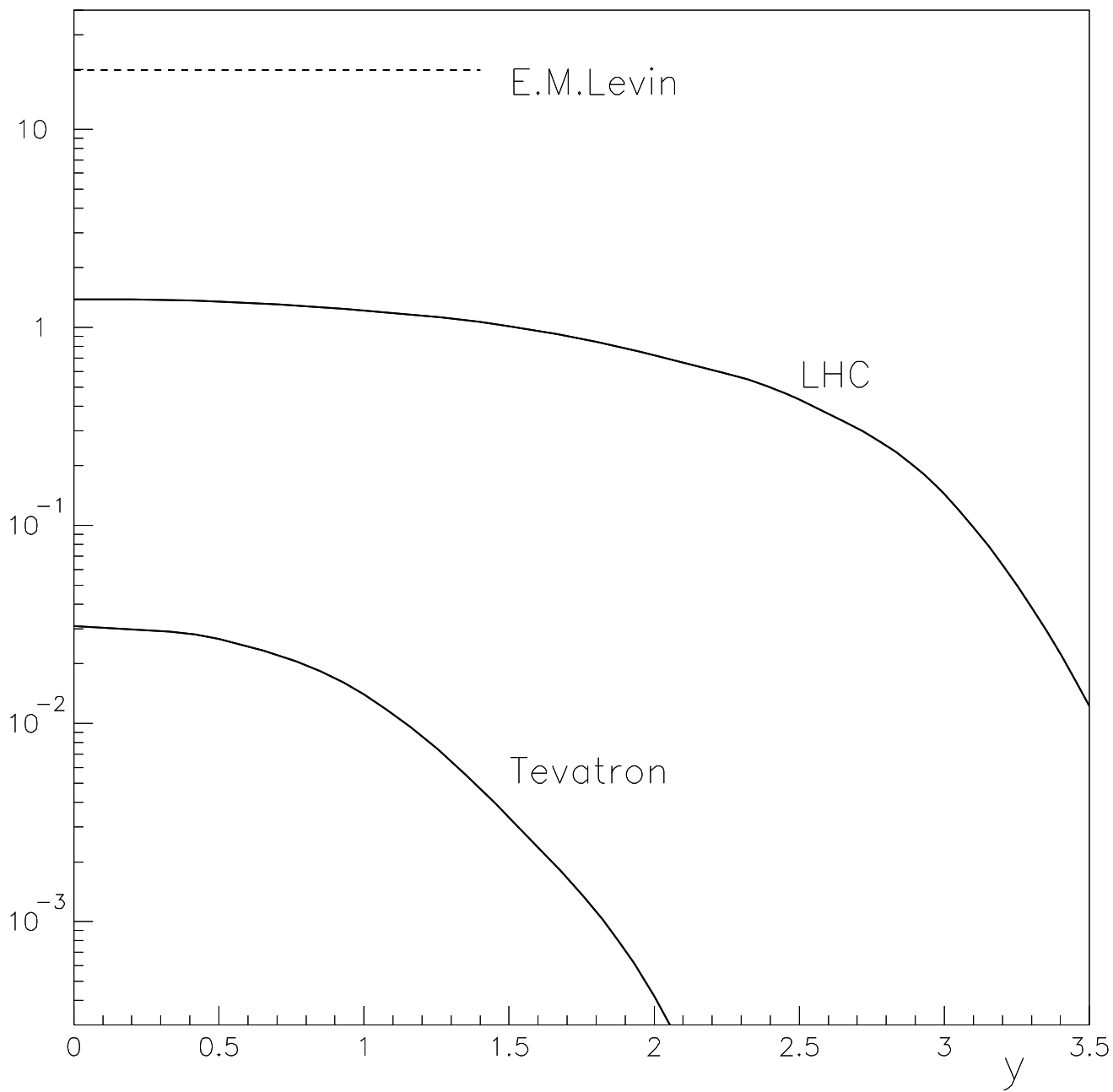
\end{document}